\documentstyle[prc,preprint,aps,epsf,array]{revtex}

\def\omit#1{}
\def\beq{\begin{equation}}
\def\eeq{\end{equation} }
\def\bea{\begin{eqnarray}}
\def\eea{\end{eqnarray}}
\def\eqref#1{Eq.~(\ref{eq:#1})}
\def\eqlab#1{\label{eq:#1}}
\def\figref#1{Fig.~(\ref{fig:#1})}
\def\figlab#1{\label{fig:#1}}
\def\tabref#1{Table~\ref{tab:#1}}

\def\secref#1{Section~\ref{sec:#1}}
\def\seclab#1{\label{sec:#1}}

\newcommand{\vslash}[1]{#1 \hspace{-0.5 em} /}
\begin{document}
\tighten

\title{Dressing the Nucleon in a Dispersion Approach}

\author{S. Kondratyuk and O. Scholten}
\address{Kernfysisch Versneller Instituut, 9747 AA Groningen, The Netherlands}
\maketitle
\vspace{1cm}
 - - - - - - - - - - - - - - - - - - - \today - - - - - - - - - - - - - - - - -

\begin{abstract}

We present a model for
dressing the nucleon propagator and vertices. In the model the use of
a K-matrix approach (unitarity) and
dispersion relations (analyticity) are combined.
The principal application of the model lies in pion-nucleon scattering
where we  discuss effects of the dressing on the phase shifts.
\omit{
We present a model for pion-nucleon scattering in which unitarity constraints
are taken into account through the use of the K-matrix formalism. In the
calculation of the K-matrix dressed vertices and propagators are employed.
The consistent dressing procedure is based on the use of dispersion integrals.
In this way analyticity constraints are taken into account.
We also discuss effects of the dressing on calculated pion-nucleon
phase shifts.
}
\end{abstract}

\bigskip
\noindent
{\bf 1999 PACS} numbers: 13.75.Gx, 11.55.Fv, 13.75.-n\\
{\bf Key Words}  Few-body systems; Pion-Nucleon Scattering;  Form Factors;
Self-Energies; K-matrix approach; Dispersion relations\\
\pacs{13.40.Gp, 21.45.+v, 24.10.Jv, 25.20.Lj}





\section{Introduction}

The properties of unitarity and analyticity have been exploited in various
theoretical approaches to pion-nucleon scattering.
Unitarity is
usually implemented either by solving a relativistic wave equation for the
scattering amplitude 
\cite{pearce,gross,julich,vlad,afnan}
or using a K-matrix formalism 
\cite{gouds,schkorpas,feusmos,korschtim}. Analytic properties, which are related
to the condition of causality, have allowed one to derive useful 
dispersion relations for various amplitudes \cite{bogol,barton}.

\omit{
Pion-nucleon scattering is a good testing ground for different theoretical
approaches to hadron physics~\cite{weinb,bjorkdrell,bogol}.
Loops and dressing and causality is the problem!!!!!!
Recently, the understanding that the pion is the
(pseudo-)Goldstone boson associated with the spontaneously broken chiral
symmetry of
QCD has allowed one to treat low-energy pion-nucleon scattering
using  effective field theory methods 
(see \cite{weinb,meissn,fettes,ellis} and references therein).
Also, a number of models
have been developed, such as (unitarized) tree-level
approaches~\cite{korschtim,schkorpas,olss,bof,gouds,feusmos,greewyc},
models  based on various relativistic equations (in the spirit of
the Bethe-Salpeter equation 
or one of its three-dimentional approximations)
\cite{pearce,julich,gross,vlad,afnan}
and others, e.~g.\
Ref.~\cite{sato}\footnote{References to relevant works can be found
in the recent paper \cite{afnan} I don't understand}.
Through the inclusion of various baryon- and meson-resonances such
models usually describe the data at pion laboratory
energies up to several hundred MeV.
}

In the present work we have developed an approach in which constraints
due to both unitarity and analyticity are incorporated 
by using both a K-matrix method~\cite{gouds,korschtim} and
dispersion relations~\cite{bogol,bincer}.
The approch used consists of two separate stages. 
In the first effective 2- and 3-point Green's functions (i.e. propagators and
vertices) are built which incorporate non-perturbative
dressing due to regular (non-pole) parts of loop diagrams.
In the second stage a K-matrix formalism is used to calculate the
T-matrix, where the kernel, the K-matrix, is constructed from tree-level
diagrams using the dressed vertices and propagators calculated in the first 
stage. Through the use of the K-matrix formalism
the pole contributions are taken into account which were left out in
the first stage.
The T-matrix obtained from thus constructed K-matrix will contain the
principal value parts of loop integrals, which is important for
implementing analyticity in the K-matrix framework.
The model as presented is geared to the calculation of pion-nucleon
scattering and includes a consistent dressing of interacting nucleons,
expanding the method of Ref.~\cite{kondsch_prc}.
Since the dressing is formulated in terms of effective vertices
and propagators through the use of form factors and
self-energies, a broader application might be possible.

In the calculation of the propagators and
vertices a technique based on the use of dispersion relations is used.
This allows us to arrange the
dressing of the nucleon such that the only $\pi N N$ vertices needed 
throughout the procedure are those with one virtual external nucleon line
(half-off-shell vertices). For the construction of the K-matrix we also need
only these $\pi N N$ vertices.
Being interacting 3-point Green's functions, such vertices
are not measurable quantities. In particular, they
depend on the representation of interpolating fields in the Lagrangian.
A wide class of field transformations leave the
S-matrix (and therefore the observables) invariant while changing
the interaction Lagrangian and hence the vertices
\cite{nels_dyson_case,haag_green_ekst,chish_kamef_coleman,scherfear}.
In this work all calculations
are performed in a particular representation in which self-energies and
vertex modifications are calculated consistently
with the K-matrix. In particular,
4- and higher-point Green's functions are absent in this representation.
In \secref{cha} we show that the field ambiguity can be
taken advantage of to change to a different representation in which the nucleon
self-energy vanishes and no 4- (or higher-) point vertices are introduced.
While leading to the same observables in virtue of the
equivalence theorem \cite{nels_dyson_case,chish_kamef_coleman}, 
this representation
is convenient for interpreting the effects of nucleon dressing in
terms of effective $\pi N N$ vertices.
Rather soft form factors in 3-point vertices are dynamically generated through
the dressing.

In the dressing
it is essential to include the low-lying meson and baryon degrees of freedom,
in particular, beside the pion, the $\rho$-- and $\sigma$-- mesons and the
$\Delta$--resonance. The associated coupling parameters were fixed by
considering phase shifts for pion-nucleon scattering. A good agreement
could be obtained for pion energies exceeding 400 MeV, adjusting only 5
parameters.
The requirement that the dressing procedure converges for a given bare
$\pi N N$ form factor puts additional
constraints on the allowed range of these parameters.

In discussing the calculated
phase shifts, we focus primarily on effects of the dressing.
These can be regarded as effects of the the explicit inclusion of
the principal value parts of loop integrals (which are
omitted in the usual approximation for the K-matrix
\cite{gouds,feusmos,korschtim}). Effects of
principal value parts were also considered in Ref.~\cite{pearce}
in a calculation based on three-dimentional reductions of the Bethe-Salpeter
equation.
In our approach we have kept the most general Lorentz structure for
the  $\pi N N$ vertex.
Another difference with~\cite{pearce} is that we obtain the principal value parts
through the use of dispersion integrals.

A general description of the model is given in \secref{mod}.
Details of the calculation of the main building blocks of the K-matrix -- 
dressed vertices and propagators -- are given in Sections III and IV,
respectively. In \secref{cha} we construct a representation which is 
convenient for interpreting results of the dressing. 
Effects of the dressing on 
calculated phase shifts for pion-nucleon scattering are discussed in 
\secref{dis}. Concluding remarks are made in \secref{con}.

\section{Outline of the model} \seclab{mod}

We start by giving the main formulae of the K-matrix approach in the context of
pion-nucleon scattering. 
The S-matrix S is expressed in terms of
the scattering amplitude $\mathcal{T}$ (the T-matrix) by  
\beq
S=1+2i{\mathcal{T}} \;.
\eqlab{s_matr}
\eeq
In principle, the T-matrix can be obtained by
solving the Bethe-Salpeter equation, 
\beq
{\mathcal{T}}\,=\,V\, +\,V\,{\mathcal G}\,{\mathcal T},
\eqlab{k1}
\eeq
where the kernel (potential) $V$ is the sum of irreducible
diagrams describing the scattering, and $\mathcal {G}$ is a dressed
$\pi N$ propagator.

In general, any integral over the 4-momenta in $\mathcal {G}$ can be
split into its pole and principal value (regular) parts,
\beq
{\mathcal G}\,=\,i\delta+\,{\mathcal G}^R, 
\eqlab{k2}
\eeq
where the pole part $i\delta$ is the contribution of the real (on-mass-shell)
$\pi N$ state, and $\mathcal {G}^R$ corresponds to the
propagation of virtual (off-shell) nucleon and pion.
According to Cutkosky rules \cite{cut},
$i\delta$ contains the imaginary parts of the invariant functions which
parametrize $\mathcal {G}$, and $\mathcal {G}^R$ contains the real parts.

It is this separation of the pole and principal value parts of
propagators that is exploited in the K-matrix formalism. Namely, on defining
the K-matrix by the equation
\beq
K\,=\,V\,+\,V\,{\mathcal G}^R\,K\,,
\eqlab{k3}
\eeq
\eqref{k1} can be written in the form
\beq
{\mathcal{T}}\,=\,K\,+\,K\,i\delta\,{\mathcal{T}}.
\eqlab{k4}
\eeq
This can be formally solved, yielding the central equation
of the K-matrix method \cite{newt}:   
\beq
{\mathcal T}= \frac{1}{1- K i\delta}\,K.
\eqlab{t_via_k}
\eeq

The S-matrix can be obtained in a two step approach:
1) given a hermitian potential $V$, calculate $K$ according to
\eqref{k3}, and 2) solve \eqref{k4} to calculate the amplitude
${\mathcal T}$ and use \eqref{s_matr} to calculate the S-matrix.
This scheme is equivalent to solving the Bethe-Salpeter equation \eqref{k1} and
as such should provide the full unitary and analytic S-matrix.
Given a hermitian K-matrix, \eqref{k4} can be solved
relatively easily (for instance, by expanding $K$ in partial waves and
using \eqref{t_via_k}), and a unitary S-matrix is
obtained. The simplicity of \eqref{k4} is due to the fact that
it involves integrals only over the mass-shells of internal particles.
The problem of solving \eqref{k3} is harder since one has to integrate
over off-shell 4-momenta of ${\mathcal G}^R$.
For this reason one usually avoids solving \eqref{k3} in K-matrix models,
setting $K=V$
\cite{gouds,schkorpas,feusmos,korschtim}.
The main drawback of this approximation is that
the principal value parts of the integrals are
completely ignored, i.\ e.\ ${\mathcal G}^R=0$, and therefore
analyticity of the amplitude cannot be fulfilled.

We take the potential $V$ as a sum of tree diagrams.
These include the s-  and u-channel diagrams with an intermediate
nucleon and $\Delta$-resonance plus the t-channel diagrams
with an intermediate $\rho$- and $\sigma$-meson.
The tree diagrams are calculated using the free propagators and bare vertices.
Note that we require that 4- and higher-point vertices do not participate
in the construction of this potential. In other words, the bare interaction
Lagrangian is assumed to contain 3-point vertices only.
According to \eqref{k3}, the K-matrix should
be obtained by dressing $V$ with the principal value parts of loop integrals.
We construct the K-matrix as
the sum of skeleton diagrams 
shown in \figref{1}. It has the
same form as $V$, except that it contains dressed  $\pi N N$ vertices and
nucleon propagators and dressed propagators of
the $\Delta$, $\rho$ and $\sigma$.

Once the K-matrix is constructed,
the T-matrix is calculated from \eqref{t_via_k} using a partial wave
decomposition as in Refs.~\cite{schkorpas,korschtim}. 
The form of the K-matrix in \figref{1}
implies that physical one-nucleon -- one-pion intermediate states are
explicitly included in the unitarization procedure. Thus,
the S-matrix obtained from \eqref{s_matr} obeys $1 N-1 \pi$ unitarity exactly.

\subsection{Dressing procedure} \seclab{mod-dre}

The calculation of the dressed $\pi N N$ vertex and the nucleon propagator
is based on a system of integral equations, shown diagrammatically
in \figref{2}. In the equation for the vertex, the external on-shell
lines for the outgoing nucleon (on the right) and the pion, as well
as the off-shell line for the incoming nucleon, are stripped away, as
indicated by dashes on these lines.
The solution is obtained in an iteration procedure and is described
in detail in Ref.~\cite{kondsch_prc}
(for the case including the nucleon and pion only).
Here we shall repeat the main points.

Every iteration step (say, step number $n$) proceeds as follows.
The imaginary or pole contributions of
the loop integrals are obtained by applying cutting rules
\cite{cut,velt} to both the propagators and the
vertices. Since the outgoing nucleon
and the pion are on-shell, the only kinematically allowed cuts for the
vertex loops are those shown by the curved lines in \figref{2}.  
In calculating these pole contributions, we retain
only real parts of the loop integrals for vertices and self-energies
from the previous step $n-1$, as dictated by \eqref{k3}.

The real parts of the form factors and self-energy functions
are calculated at every iteration step by applying
dispersion relations \cite{bogol,bincer} to the imaginary parts. For example,
for the form factors at the iteration step $n$ we have
\beq
Re\,G^n_i(p^2) = G^0_i(p^2)+ \frac{\mathcal{P}}{\pi}
      \int_{(M+\mu)^2}^{\infty} \!\! dp^{\prime 2}\,
    \frac{Im\,G^n_i(p^{\prime 2})}{p^{\prime 2}-p^2}\,,
\eqlab{dispG}
\eeq
where $i$ labels the structure of the vertex, 
pseudoscalar ($S$) or pseudovector ($V$), see \eqref{pi_non_noff_pspv}.
$M$ and $\mu$ are the nucleon and pion masses.
$G^0_i(p^2)$ are the form factors in a
bare $\pi N N$ vertex, the first term on the right-hand side of
\figref{2}.

This procedure is repeated until a converged solution is reached.
We use a normalized root-mean-square
difference $d_n$ between two subsequent iteration steps
$n$ and $n+1$ for the form factors and self-energy functions.
The convergence criterion is that $d_n < 10^{-4}$ for at least a hundred
iteration steps.
As zeroth iteration step the
bare $\pi N N$ vertex  and the free nucleon propagator are taken.
The solution of the equations in \figref{2} is equivalent to a dressing
of the potential $V$ with the principal value parts of loop integrals.


Despite the explicit use of dispersion integrals, analyticity in the model is
obeyed only ``approximately".
The violation of analyticity comes in at the level of bare form factors
needed as part of the regularization of dispersion integrals.
Strictly speaking, singularities of a bare form factor should give rise to
additional residue contributions to the dispersion relation.
To evaluate such a residue, one would have
to know the behaviour of the function to which the
dispersion relation is applied at the singularity of the bare
form factor. In general however, this behaviour is not known.
One way to circumvent this difficulty is to choose the bare form factor with
singularities that are as remote from the region of physical
interest as possible, implying a large width in general.
This is supported by the fact that the width of the form factor should be
larger than the masses of mesons included explicitly.

\omit{An exponential
bare form factor, such as the one used in this calculation, \eqref{G0},
has an essential singularity at infinity.
This invalidates the rigorous derivation of dispersion relations
because the Cauchy integral formula cannot be applied with a contour of
infinite radius.
However, given a sufficiently fast falloff of the
full dispersion integrand in the complex plane, 
the contour can be closed along a circle whose radius is
finite (not to cross the singularity), but large enough
(to approximate the integrand along the circle by zero).
Thus, the additional residues are neglected in the calculation.
and the large width of the bare form factor),
the additional residues are neglected.}

We remark that -- as a consequence of Liouville's theorem -- the
difficulty with additional singularities in the complex plane
is inherent in any approach where phenomenological
form factors are used.

\omit{
Rigorous dispersion relations could be used if one utilized a
bare form factor whose only singularities would be branch
points along the cut $p^2 > (m+m_{\pi})^2,\, Im\,p^2=0$.
Associated with these branch points would be contributions
to the imaginary parts (of the form factors or self-energy functions).
To take these additional contributions into account throughout the
iteration procedure is a non-trivial task
which entails significant comlications in applying the
cutting rules. Therefore, in the present calculation we opted for
a more tractable approach and chose the exponential bare form factor as given in
\eqref{G0}.
}

\section{Vertices} \seclab{ver}

\subsection{The $\pi N N$ vertex} \seclab{ver-nn}

The general $\pi N N$ vertex can be parametrized in terms of four
Lorentz-invariant functions (form factors) \cite{kazes}. The form
factors may depend on the 4-momenta squared of each of the three external 
lines.
For the K-matrix in this model we need $\pi N N$ vertices
with one virtual nucleon (4-momentum $p$), while the other nucleon ($p'$) and
the pion are on the mass-shell.
Such vertices are conventionally
called ``half-off-shell'' vertices. 
The general Lorentz and isospin covariant form of this vertex can be
written \cite{kazes}
\footnote{Here and throughout the paper,
we use the notation of Ref.~\cite{bjorkdrell}.}
\beq
\tau_\alpha \,\Gamma(p)=
\tau_\alpha\, P_{+}(p')\gamma^5
\left[ G_S(p^2) + P_{+}(p) G_V(p^2) \right] ,
\eqlab{pi_non_noff_pspv}
\end{equation}
for an incoming virtual nucleon.
Here $G_S(p^2)$ and $G_V(p^2)$ are pseudoscalar and pseudovector
form factors, $\tau_\alpha, \alpha=1,2,3$, are Pauli isospin matrices, and 
$P_{+}(p) \equiv (\vslash{p}+M)/(2M)$.
In the course of the dressing procedure
the most general structure \eqref{pi_non_noff_pspv} of the $\pi N N$ vertex is
maintained.

\omit{
In the following we will also need the vertex \eqref{pi_non_noff_pspv} written
in the basis of the projection operators on
positive- and negative-energy states of the off-shell nucleon,
$\Lambda_{\pm}(p) = (\pm \vslash{p}+W)/(2W)$, with the invariant mass
of the off-shell nucleon $W=\sqrt{p^2} \ge 0$,
\beq
\Gamma(p)=
P_{+}(p')\gamma^5 \left[ \Lambda_{+}(p)\,F(W) + \Lambda_{-}(p)\,F(-W) \right],
\eqlab{pi_non_noff_negpos}
\eeq
where
\beq
F(\pm W)=G_{PS}(p^2)+G_{PV}(p^2)\frac{m \pm W}{2m}.
\eqlab{f_via_pspv}
\eeq
}

The bare vertex in the dressing procedure is chosen as
\beq
G_V(p^2)=f_N\,(1-\chi) \,G^0(p^2) \;\;\; ,\; G_S(p^2)=f_N\,\chi \, G^0(p^2)
\eqlab{bare_ff}
\eeq
with
\beq
G^0(p^2)=\exp{\left[-\ln{2}\frac{(p^2-M^2)^2}{\Lambda^4_N}\right]},
\eqlab{G0}
\eeq
Here $\Lambda^2_N$ is the half-width of the bare form factor, the parameter
$\chi$ is the amount of pseudoscalar admixture in the bare vertex, and $f_N$ is
a bare coupling constant.
The latter is fixed
from the renormalization condition imposed on the dressed vertex
at the on-shell point,
\beq
\overline{u}(p^{\prime})\,\Gamma(M)\,u(p)=
\overline{u}(p^{\prime})\,\gamma^5\,g_{\pi N N}\,u(p) \;,
\eqlab{8}
\eeq
where $g_{\pi N N}$ is the physical pion-nucleon coupling constant
(for which we take the value $13.02$ \cite{korschtim}) and
$u(p)$ is the positive-energy nucleon spinor. The
renormalization conditions for the nucleon propagator are described in
\secref{pro}. Because of the coupled structure of the equations
in \figref{2}, the renormalization of the vertex and that of the
propagator are not independent of each other.

The role of the bare $\pi N N$ vertex is two-fold. On the one hand,
it serves as the driving vertex at the zeroth iteration step. On the other hand,
it is used for regularization of the dispersion integrals.
The bare vertex is supposed to encapsulate the
physics due to degrees of freedom not included explicitly in the dressing.

\omit{ where the coefficient $1.44$ is inserted for convenience,
such that$\Lambda^2_N$ is the half-width of $G^0(p^2)$, i.\ e.\
$G^0(p^2)$ reduces to its half-maximum value 1/2 at
$p^2=m^2+\Lambda^2_N$.}
\omit{
It should be remarked that throughout the paper we
write all vertices and propagators in momentum space.
I really do not like the following!!!!!
, exactly as they appear in the calculation.
In configuration space, one can write an
action functional whose integrand is an effective Lagrangian.
On applying Feynman rules, the latter should yield
these effective vertices and propagators.
Due to the presence of form factors and self-energies in momentum space,
such a Lagrangian will in general depend on the fields and their derivatives
up to infinite order \cite{pais}.
An example of an action functional that corresponds to
vertices which reduce to the half-off-shell
$\pi N N$ vertices of the general form can be found in Ref.~\cite{kondsch_phot}.
}

\subsection{Vertices with $\Delta$, $\rho$ and $\sigma$} \seclab{ver-drs}

In principle, the dressing procedure should also apply to vertices
describing the coupling to the $\Delta$-resonance and to
the $\rho$- and $\sigma$-mesons. In such an approach
 one would have to solve a system of 10 coupled equations, instead of
the system of two equations in \figref{2}. Clearly,
pursuing this is hardly feasible.

For all other vertices except $\pi N N$ we ignore the dressing and
restrict ourselves to one particular Lorentz covariant form.
With each vertex a form factor is associaled which is required for
regularization of the loop integrals.
The propagators of the $\Delta$-resonance and the
$\rho$- and $\sigma$-mesons are dressed by the standard summation of
loop insertions as discussed in \secref{pro}.

The $\sigma \pi \pi$ and $\rho \pi \pi$ vertices for a $\sigma$ or
$\rho$ meson with 4-momentum $p=q+q'$ are taken as
\bea
(\Gamma_{\rho \pi \pi})^{\,\nu}_{\,\alpha \beta \gamma} 
&=&
(\hat{e}_{\alpha \beta \gamma})\,ig_{\rho \pi \pi} F_{\rho}(p^2) \left[ k^{\nu}-
\frac{(p \cdot k)}{p^2}p^{\nu} \right] \;,
\eqlab{rho_pi_pi_vert}\\
(\Gamma_{\sigma \pi \pi})_{\,\alpha \beta} 
&=&
-i\,\frac{g_{\sigma \pi \pi}}{\mu} F_{\sigma}(p^2) \delta_{\alpha \beta}\,(q \cdot
q') \;,
\eqlab{sig_pi_pi_vert}
\eea
where $q$ and $q'$ are the 4-momenta of the pions with isospin
indices $\alpha$ and $\beta$ respectively, $k=q-q'$ and
$(\hat{e}_{\alpha \beta \gamma})=-i \epsilon_{\alpha \beta \gamma}$. 
The $\rho$-meson carries
the isospin index $\gamma$  and the Lorentz vector index $\nu$.
$g_{\rho \pi \pi}$ and $g_{\sigma \pi \pi}$ are coupling constants
(the values of all coupling constants will be given later).

\omit{ In \eqlab{rho_pi_pi_vert} and \eqref{sig_pi_pi_vert} a form factor $F$
is introduced to render a finite value for the meson self self
energies,
}
For the vertices discussed in this section a generic form factor $F_r$ is
introduced whose functional form is similar to that of the bare $\pi N N$ 
form factor given in \eqref{G0}:
\beq
F_r(p_r^2)=
\exp{\left[-\ln{2}\frac{(p_r^2-\widetilde{m}^2_r)^2-(m_r^2-\widetilde{m}^2_r)^2}
{\Lambda^4}\right]}
\;.
\eqlab{ff_res}
\eeq
normalized to unity at the on-shell point $p_r^2=m^2_r$ with a
half-width of $\Lambda^2$, the latter taken the same for all vertices
considered in this subsection.
For the $\rho \pi \pi$ and the $\sigma \pi \pi$ vertices,
$\widetilde{m}_r$,
the position of the maximum of the form factor, is set
equal to the mass of the meson, $\widetilde{m}_r=m_r$.

\omit{
Due to the use of cutting rules for the
calculation of vertex corrections and self-energies, the efficient way
of rendering finite dispersion integrals is through form facors which
depend on the momentum of the $\rho$- and $\sigma$-mesons. For this
reason the form factors in the vertices
Eqs.~(\ref{eq:rho_pi_pi_vert},\ref{eq:rho_n_n_vert},
\ref{eq:sig_pi_pi_vert},\ref{eq:sig_n_n_vert}) are chosen as
\beq
F_r(p^2)=
\exp\left(-\frac{(p^2-\widetilde{m}^2_r)^2}{1.44 \Lambda^4_r}\right)
\;,
\eqlab{ff_res}
\eeq
the same functional form as the bare $\pi N N$
form factor given in \eqref{G0}.
This form factor reaches its maximum value $1$ at $p^2=\widetilde{m}^2_r$ and
reduces to $1/2$ at $p^2=\widetilde{m}^2_r+\Lambda^2_r$.
In order to have as few parameters as possible,
we have taken $\Lambda^2_r$  the same for all
vertices. Also, $\widetilde{m}^2_r$ is chosen equal to the meson-mass
squared.
}

The ${\rho \pi \pi}$ vertex, \eqref{rho_pi_pi_vert} chosen such that it 
vanishes when contracted with the
4-momentum $p$ of the $\rho$-meson. As a consequence,
the spin-0 part of the $\rho$ propagator does not contribute to any
matrix element
(because the projection operator on the spin-0 component is
${\mathcal{P}}^0_{\mu \nu}(p)=p_\mu p_\nu/p^2$).
The apparent singularity at $p^2=0$ of the vertex \eqref{rho_pi_pi_vert}
lies outside the kinematical range covered in the calculations.
In any case, the $1/p^2$-pole behaviour could be compensated by choosing in
\eqref{rho_pi_pi_vert} a form factor with a zero at $p^2=0$.

\omit{
Because of this, an additional $p^2$ could have been absorbed is a
suitably choosen form factor (different from the present one) such
that it is similar in the region of interest as far as phenomenology
is concerned but avoids the pole.
}
The $\rho N N$ and ${\sigma N N}$ vertices are taken as
\bea
(\Gamma_{\rho N N})^{\,\nu}_{\,\gamma} 
&=& -i\,g_{\rho N N} F_{N}(p_N^2) \frac{\tau_{\gamma}}{2}
\left[ \gamma^\nu+i\,\kappa_\rho \frac{\sigma^{\nu \lambda}q_\lambda}{2M}
\right] \;,
\eqlab{rho_n_n_vert}\\
\Gamma_ {\sigma N N} 
&=& -i\,g_{\sigma N N} F_{N}(p_N^2) \;,
\eqlab{sig_n_n_vert}
\eea
where $q$ is the (incoming) momentum of the $\rho$-meson, and
$g_{\rho N N}$, $\kappa_\rho$ and $g_{\sigma N N}$ are coupling
constants. The form factor $F_{N}(p_N^2)$, where $p_N$ is the 4-momentum
of the off-shell nuceon, is given in \eqref{ff_res} with
$\widetilde{m}_N=M$, the nucleon mass.

The $\pi N \Delta$ vertex used in this calculation can  be written as 
\beq
(\Gamma_{\pi N \Delta})^{\,\nu}_{\,\alpha} 
=i\frac{g_{\pi N \Delta}}{\mu^2}\,
T_{\alpha}                                           \omit{ F_{\Delta,N} }
 \, F_{\Delta}(p^2) \,  F_{N}(p_N^2) \,
 \left[ \vslash{p}q^{\nu}-(p \cdot q)\gamma^{\nu} \right],
\eqlab{pi_n_del_vert}
\eeq
where $p$ is the (incoming) 4-momentum of the $\Delta$-resonance and
$p_N=p-q$ is the (outgoing) nucleon 4-momentum, $g_{\pi N \Delta}$ is a
coupling constant and $T_{\alpha},\,\alpha=1,2,3$, are isospin 3/2 to 1/2
transition operators. 
The form factors $F_\Delta$ and $F_N$ are taken as in \eqref{ff_res} with
$\widetilde{m}_N=M$ and $\widetilde{m}^2_{\Delta} < M_{\Delta}^2$, the mass
squared of the $\Delta$, to
obtain a reasonable description of the $P33$- phase shift in pion-nucleon
scattering (see the discussion of results below).
Indications in favour of a $\pi N \Delta$ form factor slightly assymmetric 
with respect to the $\Delta$ mass
have been also found in other works \cite{gross,feusmos}, even though
$\pi N \Delta$ vertices different to \eqref{pi_n_del_vert} have
been used there.
The dependence of the form factor in \eqref{pi_n_del_vert} on $p_N^2$ 
turns out to be neccesary to regularize
the contribution of the third loop diagram on the righ-hand side of the
equation in \figref{2}.

\omit{
The
form factor $F_{\Delta,N}$ is taken as in \eqref{ff_res}
\beq
F_{\Delta,N}\sim \, F_N(p_N^2)\cdot F_{\Delta}(p^2) \;,
\eqlab{pi_n_del_ff}
\eeq
depending on both the 4-momenta squared of the $\Delta$
($p^2$) and the nucleon ($p_N^2$) with a proportionality constant
fixed to normalize the form factor to
unity at the on-shell point $p_N^2=m^2,\;p^2=M^2$, $M$ being the mass
of the $\Delta$.
The dependence on $p_N^2$ turns out to be neccesary to regularize
the contribution of the third loop diagram on the righ-hand side of the
equation in Fig.~1.
The form factors $F_{N}$ and $F_{\Delta}$ are specified in
\eqref{ff_res} where again we used $\widetilde{m}_N=M$. \omit{  but
$\widetilde{m}_\Delta$ is considered a parameter which is adjusted to
the $\pi-N$ phase shifts.}
To obtain a reasonable description of the $P33$- phase shift in pion-nucleon
scattering, we have to take  $\widetilde{m}^2_{\Delta} < M^2$ (see the
discussion of results below).
}

The reason for the particular structure \eqref{pi_n_del_vert} 
of the $\pi N \Delta$ vertex
is that it possesses the property
$p \cdot (\Gamma_{\pi N \Delta})_{\alpha}=0$. As a consequence,
the ``sandwich" of the spin-1/2 part of the Rarita-Schwinger $\Delta$
propagator between two $\pi N \Delta$ vertices vanishes since every term in
the spin-1/2 part of the $\Delta$ propagator
is proportional to either $p_\mu$ or $p_\nu$.
Thus only the spin-3/2 part of the $\Delta$ propagator gives rise to
non-vanishing matrix elements \cite{pasc}, and it suffices to calculate only the
spin-3/2 part of the $\Delta$ self-energy.

\omit{
Finally, we remark that, in principle, the form factors for the
resonances could depend on the 4-momenta squared of all
three particles in the vertex. Since this would introduce additional
parameters in the model, we did not choose this option.
The dependence of the form factors described above is the minimal one
required for the regularization.
}

\section{Dressed propagators} \seclab{pro}

The inverse of the dressed nucleon propagator can be written as
\beq
S^{-1}(p)=\vslash{p}-M-\Sigma(p),
\eqlab{n_prop_inv}
\eeq
with the self-energy given by
\beq
\Sigma(p) = \Sigma_L(p) -(Z_2-1)(\vslash{p}-M)-Z_2\,\delta M.
\eqlab{12}
\eeq
Here $\Sigma_L(p)$ is the contribution of pion
loops,
\beq
\Sigma_L(p)=A(p^2) \vslash{p} + B(p^2) M,
\eqlab{n_se}
\eeq
parametrized by the ``self-energy functions" $A(p^2)$ and  $B(p^2)$.
The field and mass renormalization constants $Z_2$ and $\delta M$ are
fixed by requiring that
\omit{
the
expansion of $Re\,A(p^2)$ and $Re\,B(p^2)$ contain only second and
higher powers of $(p^2-M_r^2)$ \cite{velt}.
In other words 
}
the propagator have a simple pole with a unit residue at $\vslash{p}=M$.
This yields
\begin{eqnarray}
Z_2&=&\left. 1+Re\,A(M^2)+2M^2\frac{d}{d(p^2)}\left[
Re\,A(p^2)+Re\,B(p^2)\right]  \right|_{p^2=M^2},
\eqlab{del_z2} \\
\delta M&=&\frac{M}{Z_2}
\left[ Re\,A(M^2)+Re\,B(M^2) \right] \;.
\eqlab{del_dm}
\end{eqnarray}

For dressing the $\Delta$-propagator only  a
one $\pi N$ loop approximation is used.
Similar to the nucleon self-energy, the imaginary parts of the
resonance self-energy are calculated using cutting rules
and the real parts are obtained from dispersion integrals.

As explained in the previous Section, the choice of the $\pi N \Delta$ vertex
\eqref{pi_n_del_vert} allows us to retain the spin-3/2 part of the
$\Delta$ propagator only,
\beq
P_{\mu \nu}(p)=\frac{1}{\vslash{p}-M_\Delta-\Sigma_{\Delta}(p)}\,
{\mathcal{P}}^{3/2}_{\mu \nu}(p),
\eqlab{del_prop_dress}
\eeq
with the spin-3/2 projection operator \cite{pasc}
\beq
{\mathcal{P}}^{3/2}_{\mu \nu}(p)=g_{\mu \nu}-\frac{1}{3}\gamma_\mu \gamma_\nu
-\frac{1}{3p^2}(\vslash{p}\gamma_\mu p_\nu + p_\mu \gamma_\nu
\vslash{p}) \;.
\eqlab{rar_schw_proj3/2}
\eeq
Due to the elimination of the spin-1/2 components the
$\Delta$ self-energy can be parametrized by only two Lorentz invariant
functions ($A_{\Delta}(p^2)$ and $B_{\Delta}(p^2)$) instead of
10 functions which would be needed in the general case with the spin-1/2
part present. 
The structure of the self-energy is thus the same as for the nucleon,
\eqref{n_se}. The counter-term contribution to the self-energy \cite{velt}
contains the real constants $Z_2^\Delta$ and $\delta M_\Delta$ fixed by the
renormalization as described for the nucleon.
The term $\sim 1/p^2$ in \eqref{rar_schw_proj3/2} does not lead to a singularity
if the $\pi N \Delta$ vertices from \eqref{pi_n_del_vert} are used.

The meson propagators are dressed through the insertion of a
$\pi \pi$ loop as described in \secref{mod-dre}.
The pion propagator thus remains undressed.

The dressed propagator of the $\sigma$-meson has the form
\beq
D(p^2)=\frac{1}{p^2-m_\sigma^2-\Pi_\sigma(p^2)},
\eqlab{res_prop}
\eeq
where $m_\sigma$ is the physical mass of the meson and $\Pi_\sigma(p^2)$ is its
self-energy. 
The latter can be written as a sum of the loop and
counter-term contributions,
\beq
\Pi_\sigma(p^2)=\Pi_{\sigma,L}(p^2)-(Z_\sigma-1)(p^2-m_r^2)-Z_\sigma\delta
m_\sigma^2,
\eqlab{res_se}
\eeq
where $Z_\sigma$ and $\delta m_\sigma^2$ play the role of the field and mass
renormalization constants. 
These constants are fixed by requiring that the
expansion of $Re\,\Pi_\sigma(p^2)$ contain only second and higher powers
of $(p^2-m_\sigma^2)$ \cite{velt}. In other words,
\beq
\frac{1}{p^2-m_\sigma^2-Re\,\Pi_\sigma(p^2)}
\eqlab{res_renorm}
\eeq
is required to have a simple pole with a unit residue at $p^2=m_\sigma^2$.
This yields $Z_\sigma$ and $\delta m_\sigma^2$ in terms of 
$Re\,\Pi_{\sigma,L}(p^2)$:
\beq
Z_\sigma=\left.1+\frac{d}{d(p^2)}Re\,\Pi_{\sigma,L}(p^2)
\right|_{p^2=m_\sigma^2},
\eqlab{res_z}
\eeq
\beq
\delta m_\sigma^2=\frac{Re\,\Pi_{\sigma,L}(m_\sigma^2)}{Z_\sigma}.
\eqlab{res_dm2}
\eeq

Following similar arguments as for the $\Delta$,
the structure of the vertex
\eqref{rho_pi_pi_vert} has been chosen such that only the spin-1 part of the dressed
$\rho$ propagator can be retained,
\beq
D_{\mu \nu}(p)=\frac{1}{p^2-m^2_\rho-\Pi_\rho(p^2)}\,
{\mathcal{P}}^1_{\mu \nu}(p) \;,
\eqlab{rho_prop_dress}
\eeq
where
\beq
{\mathcal{P}}^1_{\mu \nu}(p)=g_{\mu \nu}-\frac{p_\mu p_\nu}{p^2} \;,
\eqlab{rho_proj1}
\eeq
is the spin-1 projection operator and $\Pi_\rho(p^2)$ is the self-energy
which has the same structure as for a scalar particle.
\omit{
For the cut dressed propagators
of the $\Delta$- , $\rho$ and $\sigma$, we use the following
expression \cite{velt}:
\beq
D^{+}(p^2)=2\,i\,\theta(p_0)\,Im\,D(p^2),
\eqlab{res_prop_cut}
\eeq
where $D(p^2)$ is the corresponding propagator (with possible vector and
spinor indices suppressed). The $\theta$-function imposes the condition 
that the energy flowing along the cut propagator is positive.
}

\section{Changing representation} \seclab{cha}

It is known that interacting Green's functions depend on the representation
of fields in the Lagrangian. There exist a
wide class of field transfomations which do not affect the asymptotic 
behaviour of the fields and hence leave the S-matrix (and
thus all observables) invariant
\cite{haag_green_ekst,chish_kamef_coleman}.
The invariance of the S-matrix under field transformations is known as
the equivalence theorem \cite{nels_dyson_case,chish_kamef_coleman}. 

In this Section we will take advantage of this irrelevance of representation
and transform the effect of the dressing of the nucleon propagator into
new $\pi N N$ vertices.
The representation constructed in this Section is an
example of the ``physical representation" discussed in Ref.~\cite{tani}.

We introduce the notation where the subscript $\Sigma$ labels the
representation where the propagator $S$ contains a non-trivial self-energy.
The new representation is defined by the two requirements:
1) the nucleon propagator must be equal to the free propagator $S^0(p)$,
2) it must be possible to construct the K-matrix as in \figref{1},
i.\ e.\ solely in terms of 2- and 3-point Green's functions.
The new $\pi N N$ vertex $\Gamma$ must thus be a solution of the equation
\beq
\Gamma(p)\,S^0(p)\,\overline{\Gamma}(p) =
\Gamma_{\Sigma}(p)\,S(p)\,\overline{\Gamma}_{\Sigma}(p) \;,
\eqlab{repres_eq}
\eeq
where only
the dependence on the internal, off-shell, 4-momentum, $p$, is indicated.
All effects of the dressing are now contained in the
difference between the new dressed and the bare vertex,
$\Gamma-\Gamma^0$.

%
%
%

The solution of \eqref{repres_eq} can be written as
\beq
\left[Re G(p^2) \pm {W \over 2 M} Re G_V(p^2) \right]=
\left[Re G_{\Sigma}(p^2) \pm {W \over 2 M} Re G_{\Sigma,V}(p^2) \right]
\sqrt{S(\pm W) \over S^0(\pm W)}
\eqlab{repres_sol}
\eeq
where $G=G_S+G_V/2$ and $W=\sqrt{p^2} \ge 0$ is the invariant mass of the
virtual nucleon. Also, $S^0(\pm W)=-1/(M \mp W)$ and
\beq
S(\pm W)=\frac{-1}{Z_2(M \mp W)-Z_2\delta M \pm Re A(p^2)W+Re B(p^2)M}
\eqlab{prop_pos_neg}
\eeq
are positive- and negative-energy parts of the free and dressed nucleon
propagators, respectively.

In the present work we consider only those solutions for the
dressed propagator
$S$ that do not have real poles in addition to the nucleon pole $W=M$. 
With this
qualification, the solution \eqref{repres_sol} for the $\pi N N$ vertex in the
new representation is well defined, since, due to the renormalization procedure,
the ratio $S(\pm W)/S^0(\pm W)$ is positive.


An extra pole at positive $W$ would
correspond to an additional asymptotic state, different to
the free nucleon. 
To take it into account properly, 
certain modifications would
be necessary of the standard renormalization of the nucleon
field \eqref{del_z2}: 
an additional field renormalization constant would probably be required
to account for the fact that a new particle species occured as a result
of the dressing \cite{weinb}. Such a study lies outside the
scope of the present work.

\section{Discussion} \seclab{dis}



The masses of the particles included in the model are 
given in \tabref{1} \cite{korschtim} and kept fixed in the calculation of
pion-nucleon phase shifts.

\omit{
as given in \tabref{1}. The value of the $\sigma$-meson mass 
was taken the same as in Ref.~\cite{korschtim}. All of the masses were fixed and
not varied in the calculation of the pion-nucleon phase shifts, described in 
what follows. 
First we remark on the choice of the bare vertex $\Gamma^0(p)$.
Restrictions on the bare vertex $\Gamma^0$ come, on the one hand, 
from the requirement of
convergence of the dressing procedure, and, on the other hand,
from the comparison of the calculated phase shifts 
(discussed in more detail in the next Section) 
with experimental data. }

An important characteristic of the bare vertex is
its half-width $\Lambda^2_N$,
see \eqref{G0}. To investigate the dependence of the dressing on $\Lambda^2_N$,
calculations have been done for two values,  $\Lambda^2_N=2$ GeV$^2$
and $\Lambda^2_N=3$ GeV$^2$, referred to as calculations
(I) and (II), respectively.
\omit{In the calculations, $\Lambda^2_N$ was not adjusted from
a comparison with the data.
To study the dependence of the dressing on this parameter,
two sets of calulations were done: with $\Lambda^2_N=2$ GeV$^2$
and with $\Lambda^2_N=3$ GeV$^2$.
In the following, these two cases will be referred to as calculations
(I) and (II), respectively.}
The requirement that a converged
solution of the dressing procedure can be obtained, without developing
additional poles
of the propagator (see \secref{cha}), puts an upper bound on
$\Lambda^2_N$. While the exact value of this limit depends also on other
parameters
of the model, it is certain that the bare form factor cannot be
arbitrarily hard. Note however that the scale introduced by the bare form
factor, which is of the order of $M^2+\Lambda_N^2$, is larger than 
the scale due to the degrees of freedom explicitly
included in the dressing. 
\omit{Note, however, that in calculations (I) and (II)
the bare form factor reduces to its half-maximum at
$p^2 \approx 2.88$ GeV$^2$ and $p^2 \approx 3.88$ GeV$^2$, respectively,
which is larger than the scale due to the degrees of freedom explicitly
included in the dressing.}
The values of the bare coupling constant $f_N$, introduced in \eqref{bare_ff},
are given in \tabref{2}, where also the values of the field and mass
renormalization constants are listed. 

We find that
a sizable pseudoscalar admixture in
the bare vertex (with $| \chi | > 0.1$ in \eqref{bare_ff})
leads to a poor description of low energy phase shifts. 
This is
intimately related to the smallness of explicit chiral-symmetry breaking.
Besides, even without resorting to phenomenology,
the range of variation of $\chi$ is severely constrained
by the requirement of convergence.
Both calculations presented in this work were done with $\chi=0.055$.

\omit{
we find the values $f_N = 11.04$ and $f_N = 10.80$ for the
calculation (I) and (II), respectively (to be compared with 
$g_{\pi N N}=13.02$).
For the masses of the $\Delta$ and $\rho$ we used the values from \cite{pdg}:
$M \equiv m_{\Delta} =1.232$ GeV, $m_{\rho}=0.77$ GeV.
The mass of the $\sigma$-meson was assumed to be
$m_{\sigma}=0.76$ GeV and not varied in the calculations.}

The values of the parameters in
the vertices for the $\Delta$-resonance and $\rho$- and $\sigma$-mesons,
Eqs.~(\ref{eq:rho_pi_pi_vert} - \ref{eq:pi_n_del_vert}), are summarized in
\tabref{3}. 
The constants $g_{\pi N \Delta}$,
$g_{\rho \pi \pi}$ and $g_{\sigma \pi \pi}$
were fixed from the decay widths of the $\Delta$, $\rho$ and $\sigma$
\cite{pdg}.
\omit{, related to the imaginary
part of their self-energies (which in turn depend on the above constants),
be within experimental bounds \cite{pdg}.}
\omit{This gives $g_{\pi N \Delta}=0.248$, $g_{\rho \pi \pi}=6.07$ and
$g_{\sigma \pi \pi}=1.88$.}
The value of half-width $\Lambda^2$, see \eqref{ff_res}, was kept fixed 
and had to be sufficiently soft to provide convergence of the dressing 
procedure.
\omit{This value was not varied in the fit to the data and set to a value
allowed by the dressing procedure.
Similar to the bare $\pi N N$ form factor, the form factor \eqref{ff_res}
Similar to the situation with the half-width $\Lambda^2_N$ of the bare
$\pi N N$ vertex, $\Lambda^2$ could not be chosen arbitrarily large.}

The coupling constants $g_{\rho N N}$, $\kappa_{\rho}$, $g_{\sigma N N}$,
as well as the parameter $\widetilde{m}^2_{\Delta}$, 
were chosen from a comparison of the calculated $\pi N$ phase shifts with 
the data, taken from \cite{arndt}. Together with $\chi$, discussed above, the
five adjustable parameters are given in the last five columns in \tabref{3}. 
It should be stressed that
only for a rather restricted range of these constants
a convergent solution of the dressing procedure could be found.

\omit{In calculation (I) the we have
$g_{\rho N N}=7.83$, $\kappa_{\rho}=0.54$, $g_{\sigma N N}=19$; and
in calculation (II) $g_{\rho N N}=8.03$,
$\kappa_{\rho}=-1.43$, $g_{\sigma N N}=18.5$.}

\omit{
To provide a good description of the P33-phase shift,
the maximum of the form factor 
\eqref{ff_res} in \eqref{pi_n_del_vert}
has to be chosen at $\widetilde{m}^2_{\Delta}<M^2$.
In particular,
in the calculation (I) $\widetilde{m}^2_{\Delta}=M^2 0.25 + m^2(1-0.25)$, and
in calculation (II) $\widetilde{m}^2_{\Delta}=M^2 0.28 + m^2(1-0.28)$.
It is interesting that indications
in favour of a slightly assymmetric $\pi N \Delta$ form factor
have been found in other works \cite{gross,feusmos}, even though
$\pi N \Delta$ vertices different to \eqref{pi_n_del_vert} have
been used there.}
\omit{
In the calculation of $\pi N$ phase shifts, the following five
parameters were adjusted: $\chi$, $g_{\rho N N}$, $\kappa_{\rho}$,
$g_{\sigma N N}$ and $\widetilde{m}^2$. As explaineded above, these parameters
are not completely free: the
dressing proicedure provides a dynamic guidance to the choice
of these parameters.}

\omit{
Using the K-matrix approach described above,
we calculated the S- and P-wave phase shifts in $\pi N$ scattering
as function of the pion kinetic energy in the laboratory frame.
The experimental data are form \cite{arndt}.
}

The phase shifts in pion-nucleon scattering are shown in Figs.~(3) and (5)
as function of the pion kinetic energy in the laboratory system,
corresponding to calculations (I) and (II), respectively.
The solid lines are the phase shifts
calculated with the dressed K-matrix as shown in \figref{1}.
The dashed lines are obtained in the approximation where $K$ is
set equal to the potential $V$, hence without taking the dressing into
account. 

\omit{In  discussing the results we are primarily interested in
the role of the dynamic dressing of the K-matrix rather than in extracting
parameters of the resonances.}

\omit{To demonstrate the dependence on the bare $\pi N N$ form factor,
two sets of calculations are shown in Figs.~3 and 5,
corresponding to the K-matrix dressed in calculations (I) and (II),
respectively.
The solid lines are the phase shifts
calculated with the dressed K-matrix as shown in Fig.~1.
The dashed lines are obtained in the approximation where $K$ is
set equal to the potential $V$. 
Thus, a comparison of the solid and dashed
lines demonstrates effects of the dressing.} 
\omit{
As explained in \secref{mod}, the T-matrix in the former calculation includes
the principal value parts of loop integrals, whereas only the pole parts
are taken into account in the latter calculation.
in the one-particle reducible diagrams contributing to the 
scattering amplitude.}
\omit{
From a theoretical point of view it is clear that an inclusion of the pole
parts alone is inadequate for calculating the full T-matrix.
At the same time,
a comparison of the solid and dashed lines in Figs.~3 and 5 with the data
seems to neither favour nor disfavour
the importance of dressing in the K-matrix approach \footnote{In this respect,
results of Ref.~\cite{pearce} seem to be similarly equivocal}.
}
 
The effect of the
dressing on the $\pi N N$ vertex, can be seen more clearly from
Figs.~(4) (calculation(I)) and (6) (calculation (II)). 
The form factors are shown as functions
of $p^2$, the invariant mass squared of the virtual nucleon.
The representation constructed
in \secref{cha} is particularly useful 
because in it the
nucleon propagator is free and the effects of nucleon
dressing are encapsulated solely in the difference $G_{V,S}(p^2)-G^0_{V,S}(p^2)$
between the dressed and bare $\pi N N$ form factors. For this reason we do not
present results for the self-energy functions.
It should be stressed that in virtue of the equivalence theorem
\cite{nels_dyson_case,chish_kamef_coleman}, either of the two
representations described in \secref{cha}
lead to identical results for the phase shifts. 
The upper and lower 
panels contain pseudovector and pseudoscalar form factors,
respectively (please note that \figref{4} and \figref{6} have
different vertical scales). 
The dotted lines are the bare form factors, see
\eqref{bare_ff}, with the constants $f_N$ and $\chi$ given in Tables II and III.  
The dashed lines are the form factors obtained after the
first iteration step (essentially, a one-loop correction to the bare vertex)
and the solid lines are the fully dressed form factors.
A comparison of the solid  and dashed lines exhibits a
non-perturbative aspect of the dressing in the sense that it goes
beyond an inclusion of few loop corrections. It can be seen that
the ratio of pseudoscalar and pseudovector form factors remains small if the
nucleon is not far off the mass-shell.
The dash-dotted curves
correspond to the form factors in the representation
where the nucleon self-energy has not been eliminated. We see that
the dressed $\pi N N$ vertex may depend significantly on the representation
chosen. 

\omit{
The dotted curve in Fig.~4 or 6 is the bare form factor $F^0(W)$ as given by
\eqref{bare_ff} with the constants $f_N$ and $\chi$ given in \tabref{2} and 
\tabref{3}.
The dashed curve shows the form factor obtained after the
first iteration step (essentially, a one-loop correction to the bare vertex)
and the solid line is the fully dressed form factor.
A comparison of the solid  and dashed lines exhibits a
non-perturbative aspect of the dressing in the sense that it goes
beyond an inclusion of few loop corrections.
The dash-dotted curve
corresponds to the form factors in the representation
where the nucleon self-energy has not been eliminated. We see that
the dressed $\pi N N$ vertex may depend significantly on the representation
chosen.}
\omit{the field transformation, effectively described by \eqref{repres_eq},
changes the dressed $\pi N N$ vertex significantly.
while leaving the physical observables the same.}

Comparing the form factors in \figref{4} with those in \figref{6}, 
we conclude that the
converged solution depends strongly on the width of the bare form factor.
However, independent of this width, the dressing causes
considerable softening of the form factor at higher invariant masses.
The results shown in Figs.~(3) and (5) suggest that
it is possible to obtain a reasonable description of phase shifts
up to pion laboratory energies of about 400 MeV starting from 
bare form factors with rather different widths.

\omit{
With only five parameters listed above, constrained by the requirement of
convergence at that, an acceptable overall description of phase shifts is
achieved up to pion laboratory energies of about 400 MeV.
This suggests that the developed dressing procedure provides a
physically reasonable method for studying higher-order correction to the
(unitarized) Born approximation traditionally adopted in the K-matrix models
\cite{korschtim,schkorpas,olss,bof,gouds,feusmos,greewyc}.}

In this model only one resonance of the $\pi N$ scattering,
the $\Delta$, was included.
The lack of other resonances becomes especially
conspicuous at higher energies.
In fact, in the calculations of phase shifts we also included
the Roper resonance, though it is not taken into account in the dressing.
This improved the calculated P11 phase shift at energies of about 300 MeV
and higher, with a negligible effect on the other phase shifts.
In principle, the Roper can be easily included in the dressing,
as well as other important degrees of freedom (for example, the S11-resonance).
In the present version of the model
we have forgone doing so, limiting ourselves to the lowest lying 
$\Delta$-resonance only.

\section{Conslusions} \seclab{con}

We have presented a model in which considerations of unitarity and anlyticity
(causality) are implemented in the K-matrix approach to pion-nucleon scattering.
The principal ingredient of the model is the dressing procedure, formulated
in terms of half-off-shell vertices and propagators, the building blocks of
the K-matrix. Analyticity properties are exploited through the use of
dispersion relations to obtain the principal value parts of loop integrals 
required for the unitarization of the scattering amplitude.   

The five parameters of the model are constrained rather much by the requirement 
of convergence of the dressing procedure. By the same token, there is 
an implicit interdependence of the parameters. 
This means that a comparison with
experiment is an important test for this approach.
We showed that a good overall description of phase shifts 
in pion-nucleon scattering can be achieved 
at the energies exceeding the scale due to the
degrees of freedom explicit in the model.
This suggests that the developed dressing procedure provides a
physically reasonable method for studying higher-order correction to the
(unitarized) Born approximation traditionally adopted in the K-matrix 
approach.

\acknowledgements

This work is part of the research program of the ``Stichting voor
Fundamenteel Onderzoek der Materie'' (FOM) with financial support
from the ``Nederlandse Organisatie voor Wetenschappelijk
Onderzoek'' (NWO). We  would like to thank
Alex Korchin, Rob Timmermans and John Tjon for discussions.

\begin{table}
\caption[1]{Physical masses of particles included in the model. All the
masses were fixed in the calculations of the $\pi N$ scattering phase phifts.}
\begin{center}
\begin{tabular}{c|c|c|c|c|c}
Particle & $N$ & $\pi$ & $\Delta$ & $\rho$ & $\sigma$ \\
\tableline
Mass, GeV& 0.938 & 0.138 & 1.232 & 0.770 & 0.760 \\
\end{tabular}
\end{center}
\label{tab:1}
\end{table}

\begin{table}
\caption[2]{Bare $\pi N N$ coupling constant and field and mass renormalization 
constants obtained in calculations (I) and (II).}
\begin{center}
\begin{tabular}{c|c|c|c|c|c|c|c|c|c}
Calculation &  $f_N$ & $Z_2$ & $Z_2^{\Delta}$ & $Z_\rho$ & $Z_\sigma$ &
$\delta M$ (GeV) & $\delta M_{\Delta}$ (GeV) & $\delta m^2_{\rho}$ (GeV$^2$) &
$\delta m^2_{\sigma}$ (GeV$^2$) \\
\tableline
(I) & 11.04 & 0.77 & 1.10 & 1.17 & 1.05 & -0.11 & -0.07 & -0.09 & -0.65 \\ 
&&&&&&&&&\\
(II)& 10.80 & 0.60 & 1.09 & 1.17 & 1.05 & -0.28 & -0.08 & -0.09 & -0.65 \\
\end{tabular}
\end{center}
\label{tab:2}
\end{table}

\begin{table}
\caption[2]{Parameters of the model used in calculations (I) and (II). 
Parameters in the last five columns only were varied in the calculations 
of the $\pi N$ scattering phase phifts.}
\begin{center}
\begin{tabular}{c|c|c|c|c|c||c|c|c|c|c}
Calculation & $\Lambda^2_N$ (GeV$^2$) & $\Lambda^2$ (GeV$^2$) &
$g_{\pi N \Delta}$ & $g_{\rho \pi \pi}$ & $g_{\sigma \pi \pi}$ & 
$\chi$ &  $g_{\rho N N}$ & $\kappa_\rho$ & $g_{\sigma N N}$ & 
$\widetilde{m}^2_\Delta$ (GeV$^2$)  \\
\tableline
(I) & 2 & 1 & 0.248 & 6.07 & 1.88 & 0.055 & 7.83 & 0.54 & 19.0 & 1.04 \\
&&&&&&&&&& \\
(II)& 3 & 1 & 0.248 & 6.07 & 1.88 & 0.055 & 8.03 & -1.43& 18.5 & 1.06 \\
\end{tabular}
\end{center}
\label{tab:3}
\end{table}

\begin{figure}
\epsfxsize 15cm
\centerline{\epsffile[-50 300 650 520]{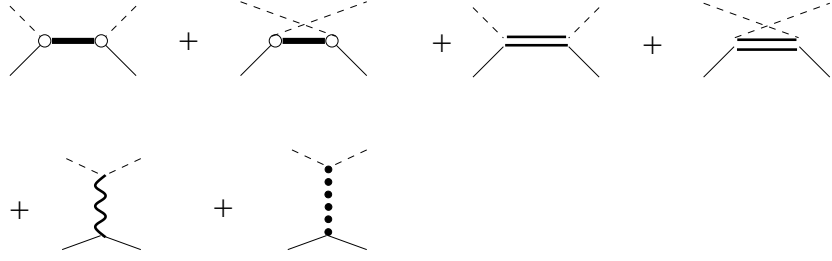}}
\caption[f1]{Diagrams included in the calculation of the K-matrix.
The solid lines are nucleons, the dashed lines pions, the solid double-lines
$\Delta$s; the wavy and dotted lines represent the $\rho$- and $\sigma$-mesons,
respectively. The intermediate propagators are dressed, as indicated by the 
thicker lines. The circle represents the dressed $\pi N N$ vertex.
\figlab{1}}
\end{figure}

\begin{figure} 
\epsfxsize 15cm
\centerline{\epsffile[-200 50 785 850]{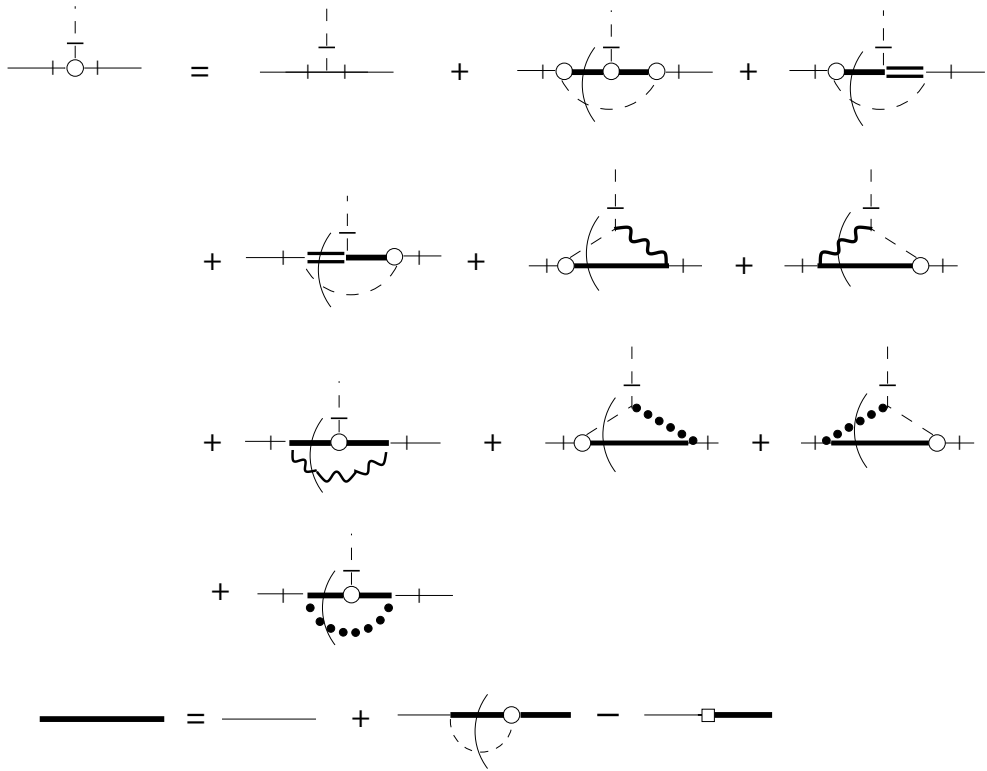}}
\caption[f2]{Diagrammatic presentation of the dressing procedure for the
half-off-shell $\pi N N$ vertex and nucleon propagator.
The notation is the same as in \figref{1}. In addition, the curved lines 
represent the cuts applied to calculate the pole 
contributions of the diagrams. The external lines are stripped away, as
indicated by the dashes. The square in the second equation is 
the counter-term contribution to the nucleon self-energy.
\figlab{2}}
\end{figure}

\begin{figure}
\epsfxsize 15cm
\centerline{\epsffile[-100 310 700 700]{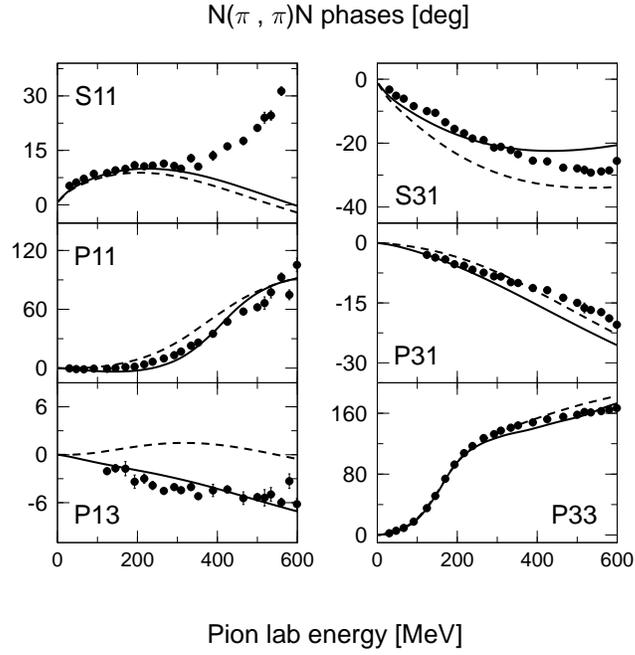}}
\caption[f3]{Pion-nucleon phase shifts from calculation (I).
The drawn curves are obtained in the full calculation. The dashed curves
represent the calculation with the bare form factors and free propagators.
The data are from \cite{arndt}.
\figlab{3}}
\end{figure}

\begin{figure}
\epsfxsize 13cm
\centerline{\epsffile[-100 120 700 860]{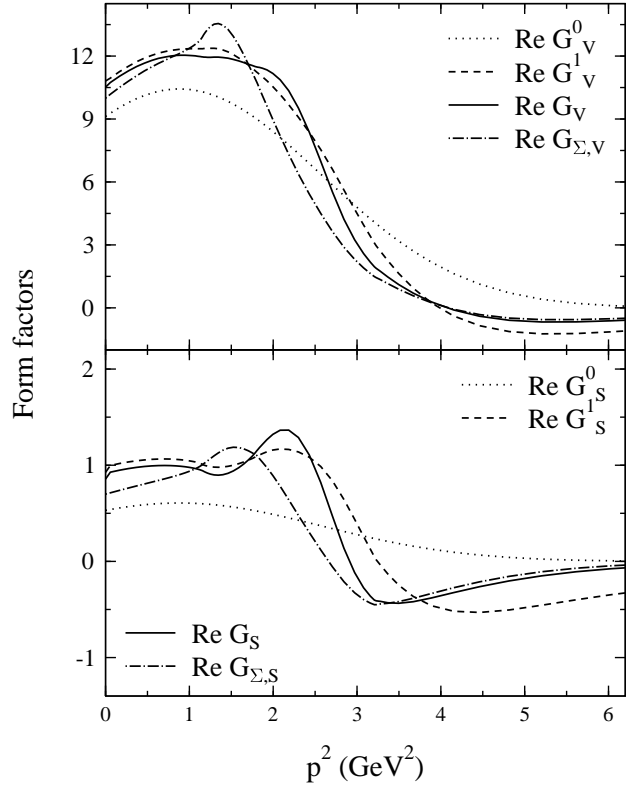}}
\caption[f4]{Pion-nucleon form factors from calculation (I). The different
curves are explained in the text.
\figlab{4}}
\end{figure}

\begin{figure}
\epsfxsize 15cm
\centerline{\epsffile[-100 310 700 700]{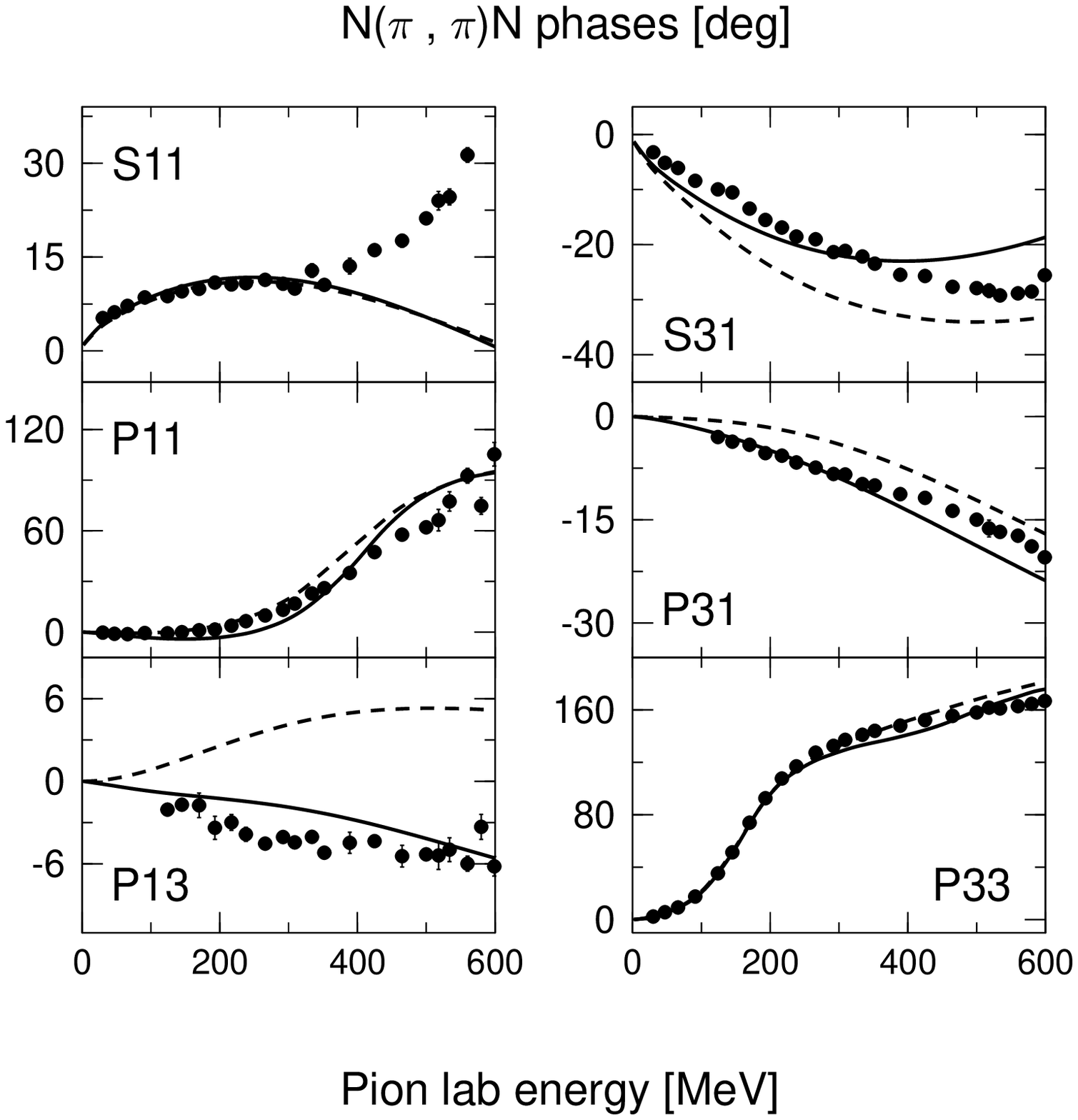}}
\caption[f5]{Same as in \figref{3}, but for calculation (II).
\figlab{5}}
\end{figure}

\begin{figure}
\epsfxsize 13cm
\centerline{\epsffile[-100 120 700 860]{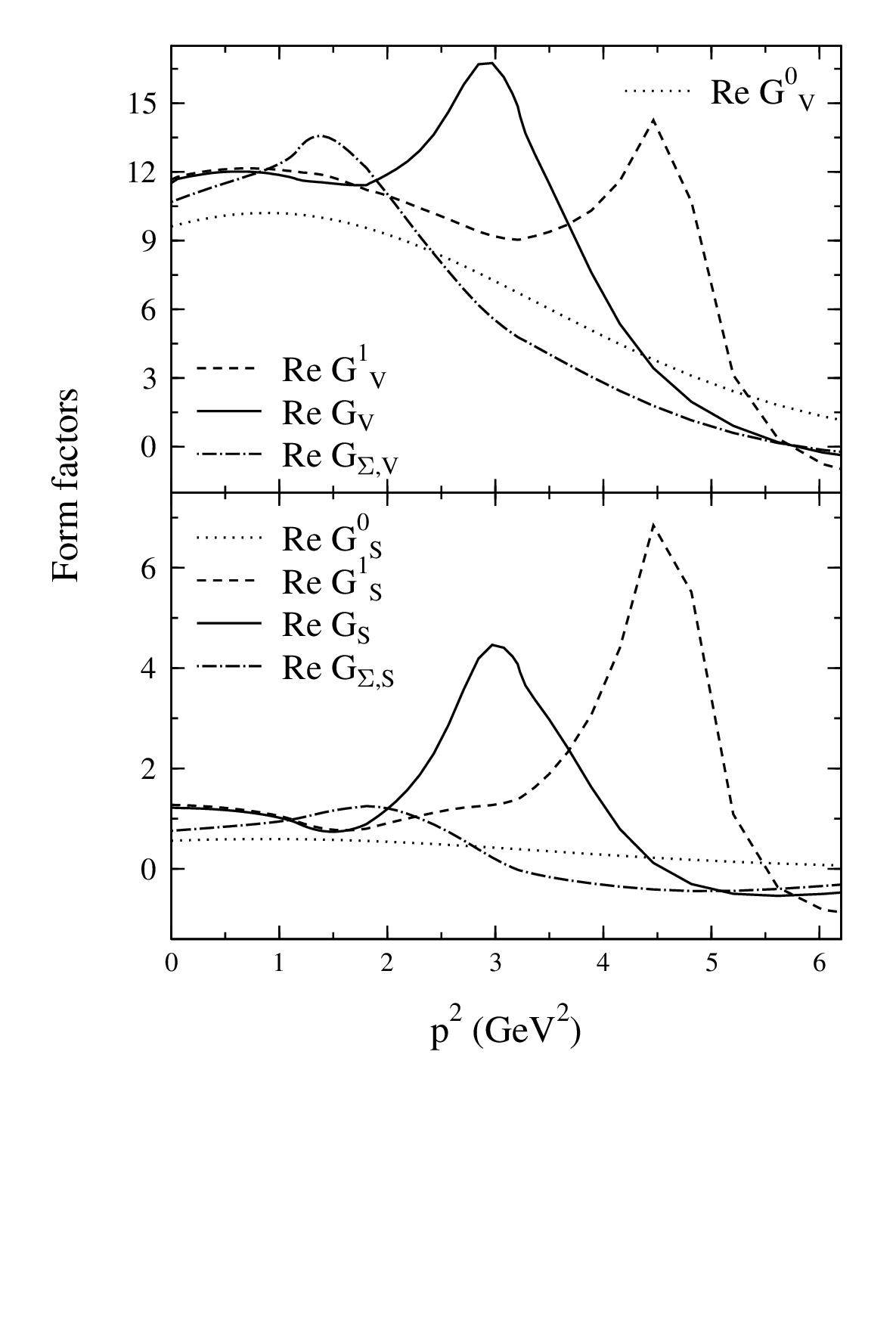}}
\caption[f6]{Same as in \figref{4}, but for calculation (II).
\figlab{6}}
\end{figure}


\begin{thebibliography}{99}



\bibitem{pearce} B. C. Pearce and B. K. Jennings, Nucl. Phys. {\bf A528}, 655
(1991).
\bibitem{gross} Franz Gross and Yohanes Surya, Phys. Rev. C {\bf 47}, 703
(1993).
\bibitem{julich} C. Sch\"utz, J. W. Durso, K. Holinde, and J. Speth,
Phys. Rev. C {\bf 49}, 2671 (1994).
\bibitem{vlad} V. Pascalutsa and J. A. Tjon, Nucl. Phys. {\bf A631}, 534c
(1998); Phys. Lett. B {\bf 435}, 245 (1998).
\bibitem{afnan} A. D. Lahiff and I. R. Afnan, Phys. Rev. C {\bf 60} (1999),
024608.
\bibitem{gouds} P. F. A. Goudsmit, H. J. Leisi, E. Matsinos, B. L. Birbrair,
and A. B. Gridnev, Nucl. Phys. {\bf A575}, 673 (1994).
\bibitem{schkorpas} O. Scholten, A. Yu. Korchin, V. Pascalutsa, and D. Van Neck,
Phys. Lett. B {\bf 384}, 13 (1996).
\bibitem{feusmos} T. Feuster and U. Mosel, Phys. Rev. C {\bf 58}, 457 (1998).
\bibitem{korschtim} A. Yu. Korchin, O. Scholten, and R. G. E. Timmermans,
Phys. Lett. B {\bf438}, 1 (1998).
\bibitem{bogol} N. N. Bogoliubov and D. V. Shirkov,
{\it Introduction to The Theory of Quantized Fields}
(Interscience Publishers, inc., New York, 1959).
\bibitem{barton} G. Barton, {\it Dispersion Techniques in Field Theory}
(W.A. Benjamin, New York, 1965).
\bibitem{bincer} A. Bincer, Phys. Rev. {\bf 118}, 855 (1960).
\bibitem{kondsch_prc} S. Kondratyuk and O. Scholten, Phys. Rev. C {\bf 59}, 1070
(1999).
\bibitem{nels_dyson_case} E.C. Nelson, Phys. Rev. {\bf 60}, 830 (1941);
F. J. Dyson, Phys. Rev. {\bf 73}, 929 (1949); K. M. Case, 
Phys. Rev. {\bf 76}, 14 (1949).
\bibitem{haag_green_ekst} R. Haag, Phys. Rev. {\bf 112}, 669 (1958);
O. Greenberg, Phys. Rev. {\bf 115} (1959); H. Ekstein, Phys. Rev. {\bf 117}, 
1590 (1960).
\bibitem{chish_kamef_coleman} J. S. R. Chisholm, Nucl. Phys. {\bf 26}, 
469 (1961); S. Kamefuchi, L. O'Raifeartaigh, and A. Salam,
Nucl. Phys. {\bf 28}, 529 (1961); S. Coleman, J. Wess and B. Zumino, Phys. Rev.
{\bf 177}, 2239 (1969).
\bibitem{scherfear} S. Scherer and H. W. Fearing, Phys. Rev. C {\bf 51}, 359
(1995);  H. W. Fearing, Phys. Rev. Lett. {\bf 81}, 758 (1998);
R. M. Davidson and G. I. Poulis, Phys. Rev. D {\bf 54}, 2228 (1996);
H. W. Fearing and S. Scherer, nucl-th/9909076.
\bibitem{newt} R. G. Newton, {\it Scattering Theory of Waves and Particles}
(Springer, New York, 1982).
\bibitem{cut}  S. Mandelstam, Phys. Rev. {\bf 115}, 1741 (1959); 
R. E. Cutkosky, J. Math. Phys. {\bf 1}, 429 (1960); G. 't Hooft and M. J. G.
Veltman, Diagrammar, CERN Yellow Report 73-09.
\bibitem{velt} M. Veltman, Physica 29, 186 (1963).
\bibitem{kazes} E. Kazes, Nuovo Cimento {\bf 13}, 1226 (1959).
\bibitem{bjorkdrell} J.D. Bjorken, S.D. Drell,
{\it Relativistic Quantum mechanics} (McGraw-Hill, 1964).
\bibitem{pasc} V. Pascalutsa, Phys. Rev. D {\bf 58}, 096002 (1998);
 Ph.D. thesis, University of Utrecht, 1998.
\bibitem{tani} Smio Tani, Phys. Rev. {\bf 115}, 711 (1959).
\bibitem{weinb} S. Weinberg, {\it The Quantum Theory of
Fields}, vol.\ 1 (Cambridge University Press, 1996).
CERN Yellow Report 73-09.
\bibitem{pdg} Particle Data Group, Eur. Phys. J. {\bf C3}, 1 (1998).
\bibitem{arndt} Virginia Tech SAID Facility, see http://clsaid.phys.vt.edu;
R. A. Arndt, I. I. Strakovskii, R. L. Workman, Phys. Rev. C {\bf 53}, 430
(1996).

\end{thebibliography}
\end{document}